\documentclass[leqno, 12pt]{article}
\usepackage{times}
\usepackage{amsmath}
\usepackage{amssymb}
\usepackage{amsfonts}

\newcommand{\hp}{\hspace*{\parindent}}
\newcommand{\B}[1]{\mbox{\boldmath $#1$}} 

\usepackage{latexsym}
\newcommand{\hem}{\hspace*{1em}}
\newcommand{\hme}{\hspace*{.5em}}

\newcommand{\hfl}{\hspace*{\fill}}
\newcommand{\nhp}{\hspace*{-\parindent}}

\newlength{\dede}     

\newcommand{\y}[1]{\mbox{$#1$}}

\newcommand{\en}{\mbox{$\in$}}

\newcommand{\lra}{\mbox{$\longrightarrow$}}
\newcommand{\Om}{\mbox{$\Omega$}}

\newcommand{\ta}{\mbox{$\tau$}}

\newcommand{\bo}{\mbox{$\bot$}}
\newcommand{\tp}{\mbox{$\top$}}
\newcommand{\se}[2]{\mbox{$#1_{#2}$}}

\newcommand{\sub}{\mbox{$\subseteq$}}

\newcommand{\Int}{\mbox{$\cap$}}

\newcommand{\Ra}{\mbox{$\Rightarrow$}}

\newcommand{\fm}[1]{\mbox{$\langle #1 \rangle$}}

\def\cuad{{\hfl \raggedleft\rule{1ex}{1ex}}}

\newcommand{\bcosq}[8]{
\begin{picture}(200,120)
\setlength{\unitlength}{.6\unitlength}
\thicklines
\put(32,20){#4}
\put(32,162){#1}
\put(39,154){\vector(0,-1){115}}
\put(65,167){\vector(1,0){100}}
\put(177,162){#2}
\put(5,90){#7}
\put(105,175){#5}
\put(177,20){#3}
\put(195,90){#8}
\put(100,8){#6}
\put(182,152){\vector(0,-1){115}}
\put(65,25){\vector(1,0){100}}
\end{picture}}

\newtheorem{theorem}{Theorem}[section]
\newtheorem{definition}[theorem]{Definition}
\newtheorem{remark}[theorem]{Remark}
\newtheorem{proposition}[theorem]{Proposition}
\newtheorem{example}[theorem]{Example}
\newtheorem{lemma}[theorem]{Lemma}
\newtheorem{notation}[theorem]{Notation}
\newtheorem{corollary}[theorem]{Corollary}

\newtheorem{fact}[theorem]{Fact}

\begin{document}
\date{}
\title{Topological representation of intuitionistic and distributive abstract logics}

\author{Andreas B.M.~Brunner\\Departamento de Matem\'atica\\Instituto de Matem\'atica\\Universidade Federal da Bahia - UFBA\\40170-110 Salvador - BA\\Brazil\\e-mail: andreas@dcc.ufba.br\and Steffen Lewitzka\\Departamento de Ci\^encia da Computa\c c\~ao\\Instituto de Matemt\'atica\\ UFBA\\40170-110 Salvador - BA\\Brazil\\e-mail: steffen@dcc.ufba.br}

\maketitle

\begin{abstract}
We continue work of our earlier paper \cite{lewbru} where \textit{abstract logics} and particularly \textit{intuitionistic abstract logics} are studied. Abstract logics can be topologized in a direct and natural way. This facilitates a topological study of classes of concrete logics whenever they are given in abstract form. Moreover, such a direct topological approach avoids the often complex algebraic and lattice-theoretic machinery usually applied to represent logics. Motivated by that point of view, we define in this paper the category of intuitionistic abstract logics with \textit{stable logic maps} as morphisms, and the category of implicative spectral spaces with \textit{spectral maps} as morphisms. We show the equivalence of these categories and conclude that the larger categories of distributive abstract logics and distributive sober spaces are equivalent, too. 
\end{abstract}

\section{Introduction}


\hp Our approach to intuitionistic and, more generally, distributive abstract logics studied in this paper is based on our previous article \cite{lewbru} where intuitionistic (and classical) logics are defined as \textit{intersection structures} (for the general notion of \textit{intersection structure} see, e.g., \cite{davpri}). All results of this paper were presented in the Brazilian Logic Conference of 2011, cf. \cite{brulewebl}. An abstract logic, viewed as an intersection structure, is essentially a system of subsets (called theories) on a set (whose elements are called formulas or expressions) such that the theories are closed under arbitrary non-empty intersections. The connectives of the underlying logic can be defined in this abstract framework by giving certain conditions that involve theories and formulas (see, e.g. Definition \ref{40} below). An advantage of this approach is that concrete logics can be translated directly into their abstract counter-parts without the explicit use of any lattice-theoretic or algebraic methods. Similar abstract views on logics have been studied over the years by several authors (see, e.g., \cite{blobro} for classical logics, and \cite{fonver} for intuitionistic logics). In fact, the name \textit{abstract logics} goes back to the seminal paper due to Brown and Suszko \cite{brosus}. In the present paper, we introduce the categories of distributive and intuitionistic abstract logics. The morphisms of these categories are \textit{logic maps} with certain additional properties. General \textit{logic maps} are discussed in \cite{lew1}; a similar concept of maps between logics was already introduced in \cite{brosus}. The notion of logic map recalls in some aspects the topological concept of a continuous map between topological spaces. In fact, it seems to be quite natural to look for a topological counterpart -- rather than a lattice-theoretical one --  of the so-defined categories of abstract logics. For this purpose, we recall some topological facts regarding sober and spectral spaces and adapt some concepts to the context of abstract logics. As the main results of this paper we are able to present duality theorems, cf. \ref{Equivalent} and \ref{EquivDistr}, showing the equivalence between the category of distributive (intuitionistic) abstract logics and the category of distributive sober (spectral) spaces with spectral maps as morphisms. \\

Topological duality results known in the literature are usually formulated for classes of certain algebras or lattices (see, e.g., \cite{bezgabkur, davpri, mir, pri, bezminmor}). The application of such results to concrete logics require a suitable process of \textit{algebraization} of the underlying logic, i.e., the establishment of a certain class of algrebras or lattices that represent the properties of the given logic. This process, which usually generalizes and extends the well-known Lindenbaum-Tarski procedere (see, e.g., \cite{blopig, jan}) is often complex and only applicable to logics which fulfill certain algebraic criteria. We believe that the process of topologizing distributive (intuitionistic) abstract logics, as described in this paper, can be extended to many other concrete logics which are given in abstract form. That is, we get a simple way to approach logics topologically avoiding the often complicated process of algebraization of a logic. \\

The paper is structured in the following manner. In the first section \ref{apprint}, we shortly recall our approach to intuitionistic abstract logics given in \cite{lewbru} which we generalize here to the class of (bounded) distributive abstract logics. In section \ref{spectralspace}, we will show an analogous result of the Boolean Prime Ideal Theorem for distributive abstract logics. Also, we define what we mean by the space of a distributive abstract logic. A series of lemmata then lead to the result that the space of a distributive logic is a sober space -- it is spectral if the logic is \textit{bounded}. This motivates our definition of \textit{(bounded) distributive space (with implication)}. We show that every spectral space is a bounded distributive space.  On the other hand, in Theorem \ref{distrspec} we establish a homeomorphism between bounded distributive spaces (with implication) and (implicative) spectral spaces. From this we derive that bounded distributive spaces are precisely the spectral spaces and that distributive spaces are sober. So we call the latter also \textit{distributive sober spaces}. The results of section \ref{spectralspace} represent a new approach to duality theorems already known and show that many intermediate logics can be dually characterized by (implicative) sober and spectral spaces, c.f. \ref{exint}. In section \ref{stabil}, we introduce {\sl stable logic maps} and present some facts necessary for the results of the last section. Stable logic maps will provide the morphisms between the objects of the category of distributive abstract logics. Finally, in section \ref{equivalence}, we define the category of \textit{intuitionistic abstract logics} $\B{IL}$ and the category of \textit{spectral spaces with implication} $\B{SI}$ and establish their categorial equivalence. If we abandon the conditions of boundedness and implication, then we get the larger categories of \textit{distributive abstract logics} and \textit{distributive sober spaces}, respectively, whose equivalence follows from the preceding results.\\

\section{Intuitionistic abstract logics}\label{apprint}

\hp Intuitionistic abstract logics, as a special case of (classical) abstract logics first studied by Bloom, Brown and Suszko \cite{brosus, blobro}, are presented as closure systems in \cite{fonver}. In \cite{lewbru} we introduce intuitionistic abstract logics as intersection structures and show the equivalence of that approach to the one given in \cite{fonver}. In this paper, we adopt the approach presented in our earlier paper \cite{lewbru} and recall in the following some basic concepts from \cite{lewbru, lew1}. 

\begin{definition}\label{20}
An abstract logic $\mathcal{L}$ is given by $\mathcal{L}=(Expr_\mathcal{L},Th_\mathcal{L},\mathcal{C}_\mathcal{L})$, where $Expr_\mathcal{L}$ is a set of expressions (or formulas) and $Th_\mathcal{L}$ is a non-empty subset of the power set of $Expr_\mathcal{L}$, called the set of theories, such that the following intersection axiom is satisfied:
\begin{equation*}
\text{If }\mathcal{T}\subseteq Th_\mathcal{L}\text{ and }\mathcal{T}\neq\varnothing,\text{ then }\bigcap\mathcal{T}\in 
Th_\mathcal{L}.
\end{equation*}
Furthermore, $\mathcal{C}_\mathcal{L}$ is a set of operations on $Expr_\mathcal{L}$, called (abstract) connectives. 
\begin{itemize}
\item We say that an abstract logic $\mathcal{L}$ is regular if $Expr_\mathcal{L}$ is not a theory, i.e., $Expr_\mathcal{L}\notin Th_\mathcal{L}$. Otherwise, $\mathcal{L}$ is singular. 
\item A subset $A\subseteq Expr_\mathcal{L}$ is called consistent if $A$ is contained in some theory $T\in Th_\mathcal{L}$. 
\item A theory $T\in Th_\mathcal{L}$ is called $\kappa$-prime ($\kappa\ge\omega$ a cardinal) if for every non-empty set $\mathcal{T}\subseteq Th_\mathcal{L}$ of size $<\kappa$, $T=\bigcap\mathcal{T}$ implies $T\in\mathcal{T}$. If $T$ is $\omega$-prime, then we say that $T$ is prime. A totally prime theory is a theory which is $\kappa$-prime for all cardinals $\kappa\le\omega$. 
\item A theory is called a maximal theory  when it is maximal in respect of set theoretic inclusion. The set of all maximal theories is denoted by $MTh_\mathcal{L}$.
\item A set of theories $\mathcal{G}\subseteq Th_\mathcal{L}$ is called a generator set if each theory is the intersection of some non-empty subset of $\mathcal{G}$. If a minimal generator set exists, then we say that $\mathcal{L}$ is minimally generated.
\item The consequence relation $\Vdash_\mathcal{L}$ is defined as follows: $A\Vdash_\mathcal{L} a :\Leftrightarrow a\in\bigcap\{T\in Th_\mathcal{L}\mid A\subseteq T\}$, for all $A\cup\{a\}\subseteq Expr_\mathcal{L}$. The consequence relation is compact if $A\Vdash_\mathcal{L} a$ implies the existence of a finite $A'\subseteq A$ such that $A'\Vdash_\mathcal{L}a$.
\item $\mathcal{L}$ is said to be compact if every inconsistent set of formulas has a finite inconsistent subset.  
\item We say that $\mathcal{L}$ is closed under chains if for any ordinal $\alpha>0$ and any chain of theories $(T_i\mid i<\alpha)$ (that is, $T_i\subseteq T_j$ for $i\le j<\alpha$), the set $\bigcup_{i<\alpha}T_i$ is a theory.
\end{itemize}
\end{definition}

Note that the notions of \textit{totally prime} theory and \textit{generator set} are very similar to the well-known order-theoretic concepts of a \textit{completely prime} element and a \textit{meet-dense} subset of a completely distribuitve lattice (see, e.g., \cite{davpri}).

\begin{fact}[\cite{lewbru}]\label{30}
Let $\mathcal{L}$ be an abstract logic. 
\begin{itemize}
\item A set of expressions $T\subseteq Expr_\mathcal{L}$ is a theory iff $T$ is consistent and closed under $\Vdash_\mathcal{L}$ (i.e. $T$ is contained in some theory, and $T\Vdash_\mathcal{L}a$ implies $a\in T$). 
\item If $\mathcal{L}$ is closed under chains, then $\mathcal{L}$ is minimally generated. 
\item $\mathcal{L}$ is closed under chains (and regular) iff the consequence relation is compact (and there is a finite inconsistent set of formulas).
\cuad
\end{itemize}   
\end{fact}

The first statement of \ref{30} follows easily from the definitions. The second statement follows from Theorem 2.11 \cite{lewbru}. The third statement follows from 2.17 \cite{lewbru}, if $\mathcal{L}$ is regular. In the singular case, it follows from basic results about closure spaces (see, e.g., \cite{davpri}).

Let $MTh_\mathcal{L},TPTh_\mathcal{L},PTh_\mathcal{L}$ denote the sets of maximal, totally prime, prime theories of logic $\mathcal{L}$, respectively. It follows that $MTh_\mathcal{L}\subseteq TPTh_\mathcal{L}\subseteq PTh_\mathcal{L}$. Furthermore, $TPTh_\mathcal{L}$ is contained in any generator set. Thus, in a minimally generated logic $\mathcal{L}$, $TPTh_\mathcal{L}$ is the minimal generator set.

The definition of \textit{intuitionistic abstract logic}, where the connectives are characterized by means of conditions over the minimal generator set, is given in \cite{lew2,lewbru}. We consider here in particular the notion of \textit{(bounded) distributive} abstract logic. 

\begin{definition}\label{40}
Let $\mathcal{L}=(Expr_\mathcal{L},Th_\mathcal{L},\mathcal{C}_\mathcal{L})$ be an abstract logic closed under chains. For a set $\{\vee,\wedge,\sim,\rightarrow\}$ of operators consider the following conditions. For all $a,b\in Expr_\mathcal{L}$ and for all $T\in TPTh_\mathcal{L}$:
\begin{enumerate}
\item $a\vee b\in T\Longleftrightarrow a\in T$ or $b\in T$
\item $a\wedge b\in T\Longleftrightarrow a\in T$ and $b\in T$
\item $\sim a\in T\Longleftrightarrow T\cup\{a\}$ is inconsistent
\item $a\rightarrow b\in T\Longleftrightarrow$ for all totally prime $T'\supseteq T$, if $a\in T'$ then $b\in T'$
\item There is a formula $\top\in Expr_\mathcal{L}$ which is contained in every (totally prime) theory (i.e. $\top$ is valid)
\item There is a formula $\bot\in Expr_\mathcal{L}$ which is contained in no (totally prime) theory (i.e. $\bot$ is inconsistent)
\end{enumerate}
If $\{\vee,\wedge\}\subseteq \mathcal{C}_\mathcal{L}$ and (i),(ii) hold, then $\mathcal{L}$ is called a distributive abstract logic. $\mathcal{L}$ is said to be bounded if in addition (v) and (vi) hold. If $\mathcal{C}_\mathcal{L}=\{\vee,\wedge,\sim,\rightarrow\}$ and (i)-(iv) hold, then $\mathcal{L}$ is an intuitionistic abstract logic. An intuitionistic abstract logic $\mathcal{L}$ with $MTh_\mathcal{L}=TPTh_\mathcal{L}$ is called a classical (or a boolean) abstract logic. 
\end{definition}

Note that an intuitionistic abstract logic is bounded.

\begin{remark}
(a) Of course, the connective of negation $\sim$ could be defined by the connectives $\bot$ and $\to$. \\ 
(b) In the literature, one may find two different ways for defining lattices. Some authors (e.g. \cite{joh}) introduce lattices as ordered sets with a greatest and a least element. Other authors refer to such lattices as bounded lattices and consider also lattices without greatest or least elements (see, e.g., \cite{davpri}). We will adopt here the latter point of view which corresponds to the situation of our abstract logics which may be bounded or not.
\end{remark}

In intuitionistic abstract logics the sets of maximal, totally prime and prime theories are in general distinct (see the discussion in \cite{lewbru}); these sets coincide in the classical case. Here comes a further example, showing this difference. 

\begin{example} \label{IntTop}
 Let $X$ be a topological space, then it is well known that the topology of $X$, denoted by $\Omega(X)$ is a frame. We have therefore the following example of an intuitionistic abstract logic. 
Let $\mathcal{L} := ( \Omega(X), Th_\mathcal{L})$ with $Th_\mathcal{L} := \{ F | \hme F$ is a filter in  $\Omega(X) \}$. Because filters are closed under union of chains, the smallest generator set are all completely irreducible filters, i.e., filters which are not intersection of other filters. Observe that in this case 
completely prime filters are completely irreducible. The connectives of disjunction and conjunction are given by $\cup,\cap$, respectively. Observe that the implication $U \to V := int (U^\mathcal{C} \cup V)$ satisfies the condition (iv) of definition \ref{40}. Negation then can be defined as $\sim U:= U\rightarrow\varnothing$. So $\mathcal{L}$ is in fact an intuitionistic abstract logic.

\nhp For $x \in X$ consider the neighborhood filter $\nu(x)$ in $\Om(X)$. This filter is completely prime or equivalently a point, cf \cite{john}. A simple calculation shows that $\nu(x)$ is not intersection of other filters in $\Om(X)$, and therefore this theory is totally prime in our abstract logic. But clearly, the neighborhood filter is in general not a maximal filter {\sl in $\Omega(X)$}, and so this theory is not maximal. Furthermore, it is not difficult to give an example of a prime filter, which is not completely prime. \cuad
\end{example}

 In \cite{lewbru} we asked for a greatest set $\mathcal{T}\subseteq Th_\mathcal{L}$ of theories such that the conditions (i)-(iv) of Definition \ref{40} remain true if we replace $TPTh_\mathcal{L}$ by $\mathcal{T}$. We call such a set \textit{the set of complete theories} $CTh_\mathcal{L}$. We have proved in \cite{lewbru} that $CTh_\mathcal{L}$ exists --- it is exactly the set of prime theories: $CTh_\mathcal{L}=PTh_\mathcal{L}$. In effect, we have shown a more general result considering appropriate notions of $\kappa$-disjunction and $\kappa$-conjuntion. Theorem 3.4 in \cite{lewbru} shows that in the presence of $\kappa$-disjunction, $CTh_\mathcal{L}$ is the set of all $\kappa$-prime theories --- this holds independently from the presence or absence of the other intuitionistic connectives.
In the case $\kappa=\omega$, this shows in particular that our notion of prime theory, introduced in an order-theoretic way, coincides with the usual notion of a prime theory $T$ in intuitionistic logic: $a\vee b\in T$ iff $a\in T$ or $b\in T$, for any formulas $a,b$. 

\begin{lemma}
A distributive abstract logic has no valid formula iff the empty set is a prime theory. On the other hand, a distributive abstract logic has no inconsistent formula iff the set of all formulas is a prime theory.
\end{lemma}

\paragraph*{Proof:}
There is no valid formula iff the intersection of all theories is the empty set iff the empty set is a theory. The empty set satisfies trivially the condition: $a\vee b\in\varnothing$ iff $a\in\varnothing$ or $b\in\varnothing$, for any formulas $a,b$. If the set of all formulas is a (prime) theory, then every formula is consistent. Now suppose that the set of all formulas is not a prime theory. Then it cannot be a theory, thus, there is an inconsistent set. Since the logic is closed under chains, it is compact (Theorem 2.14 \cite{lewbru}). That is, there is a finite inconsistent set. Its conjunction is an inconsistent formula. $\cuad$   

\section{$PTh_\mathcal{L}$ as a sober or as a spectral space}\label{spectralspace}

\hp In the following, we show that an analogue of the Boolean Prime Ideal Theorem, cf. \ref{bpi}, holds for our intuitionistic abstract logics. We define the space of the logic and show that the space of a (bounded) distributive abstract logic is a sober (a spectral) space, cf. \ref{spectral}. We introduce the notion of \textit{(bounded) distributive space} and show that spectral spaces are examples of such spaces, cf. \ref{specisdistr}. Finally, we prove that bounded distributive spaces are precisely the spectral spaces, cf. \ref{distrspec}. These theorems will primarily serve as preparations for the duality results proved in the last section. 

\begin{definition}\label{dclosure}
A set $A$ of expressions of a given distributive abstract logic is said to be closed under disjunction if $a\in A$ and $b\in A$ implies $a\vee b\in A$, for any expressions $a,b$. By $B^*$ we denote the disjunctive closure of a set $B$ of expressions, i.e. the smallest set containing $B$ being closed under disjunction. 
\end{definition}

The proof of the following analog of  the Boolean Prime Ideal theorem is standard and we sketch it.

\begin{proposition} \label{bpi}
Let $\mathcal{L}$ be a distributive abstract logic. If $T\in Th_\mathcal{L}$ and $S \subseteq Expr_\mathcal{L}$ is a non-empty set closed under disjunction such that $T \cap S = \emptyset$, then there exists a prime theory $P \in PTh_\mathcal{L}$ with $T \subseteq P$ and $P \cap S = \emptyset$.
\end{proposition}

\paragraph*{Proof:} Recall that $\mathcal{L}$ is in particular closed under union of chains and therefore $TPTh_\mathcal{L}$ is the minimal generator set and the consequence relation is finitary (see Fact \ref{30}). We will make use of Zorn's Lemma. Let $W:= \{ T'\mid T' \in Th_\mathcal{L}, \, T' \supseteq T \, \& \, T' \cap S = \emptyset \}$. Observe that $W \not = \emptyset$. Let now $\{ T_i \}_{i \in I}$ be a chain in $W$, then $\bigcup_{i \in I} T_i$ is a upper bound of $\{ T_i \}_{i \in I}$. Because our logic is closed under union of chains, $\bigcup_{i \in I} T_i$ is also a theory.

By Zorn's Lemma, there is $P \in W$ maximal. It remains to show that $P$ is prime. For this suppose that $P =T_1\cap T_2$ for any theories $T_1\supsetneq P\subsetneq T_2$. Then, by maximality, we have that $T_i \cap S \not = \emptyset$, for each $i=1,2$. Therefore, we may choose some $a_1 \in T_1\cap S$ and $a_2\in T_2\cap S$. Since each $T_i$ is the intersection of a non-empty set of totally prime theories, it follows that $a_1\vee a_2$ is contained in all these totally prime theories that generate $T_i$. Thus, $a_1\vee a_2\in T_1\cap T_2=P$. But $a_1\vee a_2\in S$, since $S$ is closed under $\vee$. Hence, $P\cap S$ cannot be empty, a contradiction. Thus, $P=T_1$ or $P=T_2$. That is, $P$ is prime. \cuad 

For the convenience of the reader, we recall basic facts concerning spectral spaces. As usual, $\overline {A}$ denotes the closure of a subset $A$ of a topological space $Y$, and $V^\mathcal{C}$ denotes the set-theoretic complement of $V$ in $Y$, i.e., $V^\mathcal{C}:= Y \smallsetminus V$. 

\begin{definition} \label{gendef} Let \y Y be a topological space, \y F \sub\
\y Y \y closed in \y Y and \y y \en\ \y Y.

\nhp a) $F$ is {\sl irreducible} iff for all closed sets
  $F_1, F_2 \subseteq Y$,
\, $(F_1 \cup F_2 = F)$ \, \Ra\ \,  $F_1 = F$ or $F_2 = F$. 

\nhp b) $y$ is a {\sl generic point for $F$} iff $ F = \overline{\{y\}}$. 

\nhp c) A  topological 
space  $Y$ is {\sl spectral} iff it satisfies the following conditions : 

$[spec\ 1]$ : \, \y Y is compact and \se{T}{0}, i.e., distinct points have distinct closures;

$[spec\ 2]$ : \, \y Y has the set of all compact opens as a {\sl basis} which is closed under finite intersections;

$[spec\ 3]$ : \, Every non-empty irreducible closed set in \y Y has a generic point. 

\end{definition}

\begin{remark}  \label{specre} a) Spectral spaces arose in Algebraic Geometry: the Zariski Spectrum
of any commutative ring with unit is spectral. In fact, the same is true
of the space of prime filters of any distributive lattice with \bo\ and \tp, cf. \cite{hoc, mir}.

\nhp b) Let \fm{Y, \tau}\ be a spectral space.  It was shown by M. Hochster in 
\cite{hoc} there is a finer topology on \y Y, \se{\tau}{c}, 
called the {\bf constructible topology}, such that \fm{Y, \se{\tau}{c}}\ is 
a Boolean space, that is, Hausdorff, compact and with a basis consisting
of clopen sets. In fact, the sets of the form \y U \Int\ $V^\mathcal{C}$, where \y U, \y V are compact opens in a basis for \fm{Y, \tau}, constitute a basis of clopens for 
\fm{Y, \se{\ta}{c}}. In particular,
every compact open in \fm{Y, \tau}\ becomes a compact clopen in 
\fm{Y, \se{\tau}{c}}. 




   
\nhp c) If \y Y, \y Z are spectral spaces, a map 
\y f : \y Y \lra\ \y Z is {\bf spectral} if it  is continuous and the
inverse image of a compact open in \y Z is a compact open in \y Y.

\nhp d) It is straightforward to check that a space is Boolean
iff it is spectral and Hausdorff.

 
\nhp e) A space with property $[spec\ 3]$, such that the generic point is uniquely determined, is also called \textbf{sober space}. Recall that sober spaces are $T_0$, but remind that sober and $T_1$ are not comparable. \cuad 
\end{remark}

The topological space of a logic is defined in the same  way as in \cite{lew1, lewbru}. Of course, within our framework of distributive logics we consider here the space of all prime theories, which have been seen are the {\sl complete} theories of these abstract logics, cf. \cite{lewbru}. 

\begin{definition}
Let $\mathcal{L}$ be a distributive abstract logic and let $X:=PTh_\mathcal{L}$. For $a\in Expr_\mathcal{L}$ we define $a^X:=\{P\in X\mid a\in P\}$. The topological space $X$ given by the base
\begin{equation*}
\Lambda(X):=\{ a^X \mid a \in Expr_\mathcal{L} \}
\end{equation*}
is called the space of the logic $\mathcal{L}$. The resulting topology is called the topology induced by $\mathcal{L}$.
\end{definition} 

\begin{proposition}
The space $X=PTh_\mathcal{L}$ of a distributive abstract logic $\mathcal{L}$ is $T_0$ and $(\Lambda(X),\cup,\cap)$ forms a distributive lattice consisting of compact open subsets of $X$. $\Lambda(X)$ contains all compact opens iff $\mathcal{L}$ has an inconsistent formula. If $\mathcal{L}$ is bounded, then $\Lambda(X)$ is a bounded lattice. \end{proposition}

\paragraph*{Proof:} The first assertions are easy to check. Note that if $\mathcal{L}$ is bounded, then in particular $\varnothing=\bot^X$ and $X=\top^X$ are basic opens. Let us show that the basic opens $a^X$, where $a \in Expr_\mathcal{L}$, are compact. For this let $a^X \subseteq \bigcup_{i \in I} b_i^X$ with $a, b_i \in Expr_\mathcal{L}$, for all $i \in I$. If $a=_\mathcal{L}\bot$ is an inconsistent expression, then $a^X=\varnothing$ and the assertion is clear. So we assume that $a$ is consistent, i.e., $a^X\neq\varnothing$. Let $B^*$ be the disjunctive closure of $B:=\{b_i\mid i\in I\}$. Recall that for any set $C$ of expressions, $C^{\Vdash_\mathcal{L}}=\{c\mid C\Vdash_\mathcal{L} c\}$. We will apply the following

{\bf Fact:} If $a^X$ has no finite covering in $\{ b_i^X\mid i \in I \}$, then $\{ a \}^{\Vdash_\mathcal{L}} \cap B^* = \emptyset$.\\
Proof of fact: Suppose $c \in  \{ a \}^{\Vdash_\mathcal{L}} \cap B^*$. Then $c$ has the form $c_1\vee...\vee c_n$, for $c_i\in B$. If $T \in X$ and $a \in T$, then $c \in T$. Recall that $CTh_\mathcal{L}=PTh_\mathcal{L}=X$ (see Theorem 3.4 of \cite{lewbru}), i.e. the prime theories are exactly the theories stable under disjunction. Thus, $a^X \subseteq (c_1\vee...\vee c_n)^X  = \bigcup\{c_i^X\mid 1\le i\le n\}$, and $a^X$ has a finite subcovering in $\{ b_i^X\mid i \in I \}$, finishing proof of fact.

Observe now that $\{ a \}^{\Vdash_\mathcal{L}}$ is consistent and deductively closed, that is, $\{ a \}^{\Vdash_\mathcal{L}}\in Th_\mathcal{L}$ (see Fact \ref{30}). Suppose $a^X$ has no finite covering in $\{ b_i^X\mid i \in I \}$. Then by the above Fact and Proposition \ref{bpi} we obtain $P\in PTh_\mathcal{L}$ with $\{ a \}^{\Vdash_\mathcal{L}} \subseteq P$ and $P \cap B^* = \emptyset$. But this is $P\in a^X$ and $P\not\in\bigcup_{i \in I} b_i^X$, contradicting the assumption that $\{ b_i^X \}_{i \in I}$ is a covering of $a^X$. Thus, $a^X$ has a finite subcovering and is compact. 

Finally, if $\Lambda(X)$ contains all compact opens, then it contains in particular the empty set. This implies the existence of an inconsistent formula, because $\varnothing=b^X$ iff $b$ is inconsistent. On the other hand, if an inconsistent formula $\bot$ exists, then $\varnothing=\bot^X\in\Lambda(X)$. Now suppose that $A\subseteq X$ is any non-empty compact open. Then there are basic opens $a_i^X$, $i\in I$, such that $A=\bigcup_{i\in I} a_i^X$. By compactness, we may assume that $I$ is finite, say $I=\{1,...,n\}$. It follows that $A=a^X$, where $a=a_1\vee...\vee a_n$. Hence, $A\in\Lambda(X)$. \cuad

\begin{corollary}
The space of a distributive abstract logic which has no valid formula is not compact. Thus, the existence of a valid formula is a sufficient and necessary condition for compactness of the space.
\end{corollary}

\paragraph*{Proof:}
Let $\mathcal{L}$ be a distributive logic with no valid formula. Then follows that $X\notin\Lambda(X)$. From the preceding Proposition it follows that $\Lambda(X)\cup\{\varnothing\}$ contains all compact opens. Thus, $X$ cannot be compact. \cuad

\begin{remark} \label{compact}
The Brouwer-Heyting intuitionistic logic generates - considering its prime theory space - a compact space, which is a spectral space. Observe that the prime theories occurring in the Brouwer-Heyting logic are the same as our prime theories, which are irreducible. This is true, because the Lindenbaum-Tarski algebra generated by an intuitionistic theory is a frame, and in particular a frame is distributive. For more details see \cite{ras}.
\end{remark}

Next we want to prove that in a distributive logic $\mathcal{L}$ every irreducible, closed non-empty set in $PTh_\mathcal{L}$ has a generic point.

\begin{proposition} \label{generic}
Let $\mathcal{L}$ be a distributive abstract logic. If $F$ is an irreducible closed non-empty set in $PTh_\mathcal{L}$, then $F$ has a generic point.
\end{proposition}

\paragraph*{Proof:} Let $F$ be an irreducible closed and non-empty set in $X:=PTh_\mathcal{L}$. We show that $P := \bigcup F$ is the generic point for $F$, i.e., $F = \overline { \{ P \}}$. Set $\Vdash := \Vdash_\mathcal{L}$ and observe that it is easy to prove that for any theories $T_1, T_2 \in PTh_\mathcal{L}$ we have 

\nhp \hfl $T_1 \in \overline{ \{ T_2 \}}$ \hem iff \hem $T_1 \subseteq T_2$. \hfl $(*)$

Observe now that 

\nhp \hfl $P \in F \hem \Longrightarrow \hem F = \overline{ \{ P \} }$. \hfl $(**)$

For this let $P \in F$, i.e., $\bigcup F \in F$. If $T\in F$ then $T \subseteq \bigcup F = P$ and by $(*)$, $T\in \overline{ \{ P \}}$.
Because $P \in F$, it is clear that $\overline {\{ P \}} \subseteq F$.

By $(*)$ and $(**)$, it suffices to prove that $P \in F$.  For this, we prove first the following 

{\bf Fact 1}: $P$ is a theory (i.e. $P$ is deductively closed and consistent).\\
Proof: First we show that $P$ is deductively closed, i.e. $P^\Vdash = P$. Let $a \in P^\Vdash$. Because $\Vdash$ is finitary, there is a finite $A \subseteq P$ with $A \Vdash a$. So there are theories $T_1, \ldots, T_k \in F$ with $a_1 \in T_1$, \ldots, $a_k \in T_k$ and $A = \{ a_1, \ldots, a_k \}$.
Observe that $a \in \bigcap \{ T\mid \hme T \in Th_\mathcal{L} \hme \& \hme a_1, \ldots, a_k \in T \}$. Because $\bigcap_{i=1}^{k} a_i^X = (a_1 \wedge \ldots \wedge a_k)^X$ we infer that \\
\hfl  $(a_1 \wedge \ldots \wedge a_k)^X \subseteq a^X$ \hfl $(***)$

Set now $b:= a_1 \wedge \ldots \wedge a_k$ and suppose that $b^X \cap F = \emptyset$. Then $(b^X \cap F)^\mathcal{C} = PTh_\mathcal{L}$. But this is $F \cap \bigcup_{i=1}^k (a_i^X)^\mathcal{C} = F$ and so, $\bigcup_{i=1}^k F\cap (a_i^X)^\mathcal{C} =F$, where the $F\cap (a_i^X)^\mathcal{C}$ are closed sets. But $F$ is an irreducible
closed set and so there exists some $j \in \{1, \ldots, k \}$ with $F = F \cap (a_j^X)^\mathcal{C}$. But then, 

\nhp \hfl $F \cap a_j^X = F \cap (a_j^X)^\mathcal{C} \cap a_j^X = \emptyset$, \hfl

\nhp and this is a contradiction, because $T_j \in F\cap a_j^X$. So, we must have $b^X \cap F \not = \emptyset$. By $(***)$ we infer that $a^X \cap F \not = \emptyset$. Therefore, there exists a $T \in PTh_\mathcal{L}$ with $a\in T$ and $T \in F$, i.e., $a \in P$ and we have proved that  $P^\Vdash = P$. 

It remains to show that $P$ is consistent. If $\mathcal{L}$ is singular, then every set of expressions is consistent. So we may assume that $\mathcal{L}$ is regular. In this case, consistency of $P$ is equivalent with the condition $P\neq Expr_\mathcal{L}$ (recall that $P$ is deductively closed). Theorem 2.17 in \cite{lewbru} yields the existence of a finite inconsistent set from which the existence of an inconsistent formula $\bot$ follows. Now the assumption $P=Expr_\mathcal{L}$ leads to the contradiction $\bot\in T$ for some prime theory $T\in F$. Thus, $P\subsetneq Expr_\mathcal{L}$, that is, $P$ is consistent. We have proved Fact 1.

We prove now the following

{\bf Fact 2}: $P$ is prime.\\
Proof: Suppose $P$ is not prime. Then there are theories $T_1,T_2$ such that $P=T_1\cap T_2$ and $T_1\neq P\neq T_1$. We choose $a\in T_1\smallsetminus P$ and $b\in T_2\smallsetminus P$. Since $T_1$ and $T_2$ are intersections of sets of totally prime theories, we get $a\vee b\in T_1\cap T_2=P$. Thus, there is some prime theory $T\in F$ such that $a\vee b\in T$, and therefore $a\in P$ or $b\in P$, a contradiction. Hence, $P$ is prime. 

It remains to show that $P \in F$. For this, let $a \in P$, then there is $T \in F$ with $a \in T$ and so $T \in a^X$. Let now $U$ be an open neighborhood of $P$, then $U \cap F \not = \emptyset$. Therefore, $P \in \overline{F} = F$. We have now a generic point $P = \bigcup F$ of the irreducible non-empty theory $F$, finishing our proof. \cuad  

The following theorem summarizes the preceding results:

\begin{theorem} \label{spectral}
Let $\mathcal{L}$ be a distributive abstract logic. Then the space $X=PTh_\mathcal{L}$ with the lattice $\Lambda(X)$ as base is a sober space. $\Lambda(X)\cup\{\varnothing\}$ contains all compact opens. $X\in\Lambda(X)$ iff $\mathcal{L}$ has a valid formula. $\varnothing\in\Lambda(X)$ iff $\mathcal{L}$ has an inconsistent formula. If $\mathcal{L}$ is a bounded distributive logic, then the space $X$ is spectral and, obviously, $\Lambda(X)$ is a bounded lattice. \cuad
\end{theorem}

In the following, we want to give some examples of spectral spaces and intuitionistic abstract and distributive abstract logics - showing that our following duality theorems hold for a great variety of logics. 

\begin{example} \label{exint}
(a) Let $\mathcal{L}$ be the Brouwer-Heyting intuitionistic logic, then we can prove  that the space generated by the intuitionistic prime theories is a spectral space.

(b) In an analog way as in example \ref{IntTop}, we see that if $\Omega$ is a frame, that is a $[\wedge, \bigvee]$-lattice -- the $\wedge, \bigvee$ distributive law holds -- then all filters in $\Omega$ as theories define an intuitionistic abstract logic. The details are similar as in the earlier mentioned example \ref{IntTop}. Remark only that every frame admits an implication $\to$ satisfying the adjunction property in an Heyting algebra, i.e., \\
\nhp \hfl $\forall x, y, z \in \Omega, \hem \hem z \leq x \to y$ \hem iff \hem $z \wedge x \leq y$. \hfl 

(c) Knowing that every Kripke frame $\underline{P}$, i.e., $\underline{P} :=(P; \leq)$ a poset,  in a Kripke model $\mathcal{K}:= (\underline{P}; \Vdash)$, gives rise to a Heyting algebra by setting $\Omega := \{ A \subseteq P | \hem A = \uparrow A$\footnote{For the definition of the up set $\uparrow A$ see the comments following this definition.}$\}$ with the inclusion order, cf. \cite{fit}, we have a lot of new examples of intuitionistic abstract logics. Remark that we have as the connectives $\wedge$ and $\vee$, simply intersection and union, respectively.  The implication is given for $A, B \in \Omega$, by $A \to B := \{ t \in \bigcup P| \hem (\uparrow t) \cap A \subseteq B \}$.

(d) Let $L_I$ be the intuitionistic Brouwer-Heyting logic. Then it is imediate that \\ 
$\mathcal{L} := (Form(L_I), Th(L_I), \mathcal{C}_\mathcal{L})$, with $Form(L_I)$ be the set of all $L_I$-formulas and $Th(L_I)$ the set of all intuitionistic theories and $\mathcal{C}_\mathcal{L}=\{\vee,\wedge,\sim,\rightarrow\}$ the usual connectives, is an example of an intuitionistic abstract logic with smallest generator set the completely prime (i.e., completely irreducible) theories. 

(e) Let $LC$ be the G\"odel-Dummett logic, given by the axiomatics $Int$ for intuitionistic propositional logic with the additional axiom scheme, $((p \to q)\vee(q\to p))$. The Kripke model for this logic is given by {\sl strongly connected} Kripke-frames $(P; \leq)$, i.e, $\leq$ is a partial order such that for all $a, b, c \in P$, if $a \leq b$ and $a \leq c$, then $b \leq c$ or $c \leq b$. We know also that the G\"odel-Dummett logic is exactly that logic which is satisfied in linearly ordered Heyting algebras, as for example $[0;1]$. Considering $Form(LC)$  the set of all $LC$-formulas and $Th(LC)$ the set of all intuitionistic intermediate G\"odel-Dummett theories, and define the connectives $\mathcal{C}_\mathcal{L}=\{\vee,\wedge,\sim,\rightarrow\}$ as in intuitionistic logic. Then $\mathcal{L} := (Form(LC), Th(LC), \mathcal{C}_\mathcal{L} )$ is also an example of an intuitionistic abstract logic with smallest generator set the completely prime (i.e., completely irreducible) theories. 


(f) Also  some other intermediate logics, as for example, the Kreisel-Putnam logic $KP$, the Jankov logic $Jn$, the Scott logic $St$, and the Anti-Scott logic $ASt$, the  Medvedev logic $Medved$, cf. \cite{fio}, etc. can be formalized within the context of intuitionistic abstract logics - by the same manner as explained in the last example (e). 


(g)  Let $L_J$ be the Johansson logic, also known as minimal logic, cf. \cite{ras}. Then we have that $\mathcal{L} := (Form(L_J), Th(L_J), \mathcal{C'}_\mathcal{L} )$, with $Form(L_J)$ be the set of all $L_J$-formulas and $Th(L_J)$ the set of all intuitionistic minimal theories and $\mathcal{C'}_\mathcal{L} =\{\vee,\wedge, \to \}$, is an example of an intuitionistic abstract logic without (intuitionistic) negation and with smallest generator set the completely prime (i.e., completely irreducible) theories. Remark that this logic, not only rejects the {\sl tertium non datur}, but also {\sl ex falso sequitur quodlibet}. Thus, the Johansson logic is an example for an intuitionstic paraconsistent logic. Although, the Johansson logic has the connective $\sim$ -- which is generally defined by $ \to \bot$ --, that connective  {\sl does not} fulfill the condition (iii) in \ref{40}, because of the paraconsistent character of this logic. 

(h) In the same manner, we can treat $L_P$ the positive logic with semi-negation, cf. \cite{ras}. Let $Form(L_P)$ be the set of all $L_P$-formulas and $Th(L_P)$ the set of all intuitionistic positive theories. Then $\mathcal{L} := (Form(L_P), Th(L_P), \mathcal{C'}_\mathcal{L} )$ is also an example of an intuitionistic abstract logic without (intuitionistic) negation and with smallest generator set, the completely prime (i.e., completely irreducible) theories. \cuad
\end{example}

Recall that if $(X,\le)$ is a partial order and $U\subseteq X$, then $\uparrow U$ denotes the set $\{y\in X\mid x\le y$ for some $x\in U\}$. As usual, we write $\uparrow x$ instead of $\uparrow\{x\}$. $U$ is called an upset if $\uparrow U=U$. Also recall that the \textbf{specialization pre-order} $\le$ on a topological space $X$ is given by $x\le y$ iff $\overline{ \{ x \} }\subseteq\overline{\{ y \}}$ iff $y$ is contained in any (basic) open which contains $x$. This pre-order is anti-symmetric (i.e. is an partial order) iff the underlying space is $T_0$.

For a topological space $X$ we denote by $\Omega(X)$ the complete lattice of open sets. In the following we assume that $X$ has a base $\Lambda(X)$ such that $(\Lambda(X),\cup,\cap)$ is a lattice and $\Lambda(X)\cup\{\varnothing\}$ contains all compact open subsets of $X$. For each $x\in X$ let $x^{\Omega(X)}=\{U\in\Omega(X)\mid x\in U\}$ and $x^{\Lambda(X)}=\{U\in\Lambda(X)\mid x\in U\}$, and finally $X^{\Omega(X)}=\{x^{\Omega(X)}\mid x\in X\}$ and $X^{\Lambda(X)}=\{x^{\Lambda(X)}\mid x\in X\}$. If $X$ is a sober space and $\le$ is its specialization order, then follows that $(X,\le)$ and $(X^{\Omega(X)},\subseteq)$ are order-isomorphic via $x\mapsto x^{\Omega(X)}$. Clearly, the sets $x^{\Omega(X)}$ are completely prime filters on the lattice $\Omega(X)$. The condition of sobriety of $X$ is equivalent with the existence of a bijection between the points and the completely prime filters on $\Omega(X)$ (see, e.g., \cite{joh}). So if $X$ is a sober space, then $X^{\Omega(X)}$ is the set of all completely prime filters on $\Omega(X)$. These facts are well-known. In the following we draw our attention to the set of prime filters on $\Lambda(X)$. 

\begin{definition}\label{ordspec}
Let $X$ be a $T_0$-space with a base $\Lambda(X)$ of compact opens such that the following hold: 
\begin{enumerate}
\item $\Lambda(X)\cup\{\varnothing\}$ contains all compact opens.
\item $(\Lambda(X),\cup,\cap)$ is a lattice.
\item Every prime filter $P$ on the lattice $\Lambda(X)$ is of the form $P=x^{\Lambda(X)}=\{U\in\Lambda(X)\mid x\in U\}$, for some $x\in X$. That is, $X^{\Lambda(X)}=\{x^{\Lambda(X)}\mid x\in X\}$ is the set of all prime filters on the lattice $\Lambda(X)$.
\end{enumerate}
We call $X$ a distributive space. A distributive space $X$ is called bounded if $\varnothing\in\Lambda(X)$ and $X\in\Lambda(X)$. --- Let $\le$ be the specialization order on the distributive space $X$. If for any two basic opens $U,V\in\Lambda(X)$, the set $U\rightarrow V:=\{x\in X\mid\forall y\ge x:\text{ if }y\in U,\text{ then }y\in V\}=\{x\in X\mid (\uparrow x)\cap U\subseteq V\}$ is a basic open, i.e. $U\rightarrow V$ is an element of $\Lambda(X)$, then $X$ is called a distributive space with implication (or an implicative distributive space).
\end{definition}

\begin{lemma}
Let $X$ be a distributive space with specialization order $\le$.
\begin{enumerate}
\item $(X,\le)$ is order-isomorphic with $(X^{\Lambda(X)},\subseteq)$ via $x\mapsto x^{\Lambda(X)}$. 
\item Every non-empty chain w.r.t. $\le$ has a supremum in $X$. Thus, $(X,\le)$ is a dcpo. 
\end{enumerate}
\end{lemma}

\paragraph*{Proof.}
(i) follows easily from the fact that $X$ is $T_0$. Let us prove (ii). Let $C=(x_i^{\Lambda(X)}\mid i\in I)$ be a non-empty chain w.r.t. $\subseteq$. Since the elements of $X^{\Lambda(X)}$ are prime filters on $(\Lambda(X),\subseteq)$, the union of $C$ is again a prime filter. Condition (iii) of the previous Definition states that this prime filter must be of the form $y^{\Lambda(X)}$ for some $y\in X$. Now (ii)  follows from the order-isomorphism $x\mapsto x^{\Lambda(X)}$ between $(X,\le)$ and $(X^{\Lambda(X)},\subseteq)$. \cuad

The following facts are well-known or easy to prove.

\begin{remark}
\begin{itemize}
\item If $X$ is any topological space with basis $\Lambda(X)$ and a order $\le$ such that $(X,\le)$ is order-isomorphic with $(X^{\Lambda(X)},\subseteq)$, then $X$ is $T_0$ and $\le$ is the specialization order.
\item In any $T_0$-space the (basic) opens are upsets with respect to the specialization order. On the other hand, if $X$ is any $T_0$-space in which every basic open is an upset with respect to a given order $\le$, then $\le$ is the specialization order.

\item Continuous maps are monotonous on the specialization order.

\item In a distributive space with implication holds adjunction. That is, for $U, V, W \in \Lambda(X)$: \\ 
\nhp \hfl $W \subseteq U \to V$ \hem iff \hem $W \cap U \subseteq V$. \hfl

\end{itemize}
\end{remark} 

The next result essentially says that in a spectral space $X$ the points are not only in bijection with the completely prime filters on $\Omega(X)$ but also with the prime filters on $\Lambda(X)$.

\begin{proposition}\label{specisdistr}
Every spectral space is a bounded distributive space.
\end{proposition}

\paragraph*{Proof.}
Let $X$ be a spectral space. By definition, the set $\Lambda(X)$ of all compact opens is a base and it forms a bounded lattice. We show that this together with sobriety of $X$ implies that each prime filter on the lattice $\Lambda(X)$ is of the form $x^{\Lambda(X)}$, for some $x\in X$. So let $P$ be a prime filter on $\Lambda(X)$. Define $G := \{ U \in \Omega(X) | \hme \exists V \in P, \, V \subseteq U \}$ to be the filter generated by $P$ in $\Om(X)$. Then we prove the following 

{\bf Fact 1}: $G$ is a completely prime filter in $\Om(X)$.\\
Proof of the fact: Let $S \subseteq \Om(X)$ such that $\bigcup S \in G$. By definition of $G$, there is $V \in P$ with $V \subseteq \bigcup S$. Observe that for all $U \in S$, $U = \bigcup_{k \in I_U} W_k$, with $W_k \in \Lambda(X)$. So $V \subseteq \bigcup_{U \in S} \bigcup_{k \in I_U} W_k$. Put $I := \bigcup_{U \in S} I_U$ (we may assume that the $I_U$ are pairwise disjoint). By compactness of $V$ there exist $k_1, \ldots, k_n \in I$ such that $V \subseteq \bigcup_{i=1}^{n} W_{k_i} \in P$. Because $P$ is prime we have that $W_{k_i} \in P$ for some $i \in \{ 1, \ldots, n \}$. Let $U \in S$ such that $W_{k_i} \subseteq U$. Then $U \in G$, showing that $G$ is completely prime.

By Fact 1 and sobriety of $X$, there exists $x \in X$ such that $x^{\Omega(X)} = G$. Therefore, $P = G \cap \Lambda(X) = x^{\Lambda(X)}$. Since the space is $T_0$, we have a bijection between the points and the prime filters on $\Lambda(X)$. \cuad

The preceding result together with the next one imply that bounded distributive spaces are exactly the spectral spaces. The proof of the following result will be useful to derive the desired equivalence between spectral spaces and intuitionistic abstract logics.

\begin{theorem}\label{distrspec}
A bounded distributive space $X$ (with implication) is homeomorphic to the (implicative) spectral space $X^{\Lambda(X)}$ with base $\Lambda(X^{\Lambda(X)})$ via the homeomorphism $x\mapsto x^{\Lambda(X)}$.
\end{theorem}

\paragraph*{Proof.}
Let $X$ be a bounded distributive space with implication. We define 
\begin{equation*}
\mathcal{L}:=(\Lambda(X),Th_\mathcal{L},\{\cup,\cap,\rightarrow,\sim\}),
\end{equation*}
where $Th_\mathcal{L}:=\{\bigcap A\mid A\subseteq X^{\Lambda(X)}$ and $A\neq\varnothing\}$, $\rightarrow$ is the implication of the space $X$, and $\sim U:=U\rightarrow\varnothing$ for any $U\in\Lambda(X)$. Note that $Th_\mathcal{L}$ is closed under intersections of non-empty subsets. Thus, $\mathcal{L}$ is an abstract logic. By definition, $X^{\Lambda(X)}$ is a generator set. By the preceding Lemma, this generator set is closed under union of chains. By Fact \ref{30}, $\mathcal{L}$ is minimally generated and its consequence relation is compact. Since $X^{\Lambda(X)}$ is exactly the set of prime filters on $\Lambda(X)$, we get $PTh_\mathcal{L}=X^{\Lambda(X)}$. $X^{\Lambda(X)}=PTh_\mathcal{L}$ contains in particular all totally prime theories (i.e., the completely prime filters on $\Lambda(X)$). The logic is bounded, since $\varnothing$ is the inconsistent formula and $X$ is the valid formula. It is clear that $\cap,\cup$ are the intuitionistic connectives of conjunction and disjunction, respectively. Let us show that $\rightarrow$ is intuitionistic implication. For this suppose $x^{\Lambda(X)}\in X^{\Lambda(X)}$ is a totally prime theory. Then $U\rightarrow V\in x^{\Lambda(X)}$ iff $x\in U\rightarrow V=\{y\in X\mid\uparrow y\cap U\subseteq V\}$ iff for all $z\ge x$: $z\in U$ implies $z\in V$ iff for all $z^{\Lambda(X)} \supseteq x^{\Lambda(X)}$: $U\in z^{\Lambda(X)}$ implies $V\in z^{\Lambda(X)}$ iff for all totally prime $z^{\Lambda(X)}\supseteq x^{\Lambda(X)}$: $U\in z^{\Lambda(X)}$ implies $V\in z^{\Lambda(X)}$. Thus, $\rightarrow$ satisfies the definition of intuitionistic implication. Now one easily checks that $\sim$ satisfies the condition of intuitionistic negation. 

In \ref{spectral} we have seen that $X^{\Lambda(X)}=PTh_\mathcal{L}$ is a spectral space with basis $\Lambda(X^{\Lambda(X)})=\{U^{X^{\Lambda(X)}}\mid U\in\Lambda(X)\}$ of all compact opens, where $U^{X^{\Lambda(X)}}=\{x^{\Lambda(X)}\in X^{\Lambda(X)}\mid U\in x^{\Lambda(X)}\}=\{x^{\Lambda(X)}\in X^{\Lambda(X)}\mid x\in U\}$. Since $X$ is a distributive space, $h:X\rightarrow X^{\Lambda(X)}$ defined by $x\mapsto x^{\Lambda(X)}$ is by hypothesis a bijection. Let $U\in\Lambda(X)$. Then $h(U)=\{h(x)\mid x\in U\}=\{x^{\Lambda(X)}\mid x\in U\}=U^{X^{\Lambda(X)}}\in\Lambda(X^{\Lambda(X)})$. Hence, $h$ is open. Now let $V^{X^{\Lambda(X)}}\in\Lambda(X^{\Lambda(X)})$. Then $h^{-1}(V^{X^{\Lambda(X)}})=h^{-1}(\{x^{\Lambda(X)}\mid x\in V\})=\{x\mid x\in V\}=V\in\Lambda(X)$. Hence, $h$ is continuous. This shows that the space $X$ and the spectral space $X^{\Lambda(X)}$ are homeomorphic via $x\mapsto x^{\Lambda(X)}$. The existence of an implication in the spectral space $X^{\Lambda(X)}=PTh_
 \mathcal{L}$ now follows from the existence of an implication in the homeomorphic space $X$. In view of the following Corollary \ref{charspec} we give an alternative proof deriving the implication in $X^{\Lambda(X)}$ from the implication in the logic $\mathcal{L}$. Note that the set $Th_\mathcal{L}$ of all theories of $\mathcal{L}$ is stable under the connective of implication. This is shown in Theorem 3.4 of \cite{lewbru}. In particular, the set of all prime theories is stable under implication. That is, we may replace the totally prime theories by prime theories in the defining condition of implication. So for $a, b \in Expr_\mathcal{L}=\Lambda(X)$ we may argue as follows: $(a\rightarrow b)^{PTh_\mathcal{L}}=\{P\in PTh_\mathcal{L}\mid a\rightarrow b\in P\}=\{P\in PTh_\mathcal{L}\mid\text{ for all prime }P'\supseteq P,\text{ if }a\in P',\text{ then }b\in P'\}=\{P\in PTh_\mathcal{L}\mid (\uparrow P)\cap a^{PTh_\mathcal{L}}\subseteq b^{PTh_\mathcal{L}}\}=a^{PTh_\mathcal{L}}\rightarrow b^{PTh_\mathcal{L}}\in\Lambda(PTh_\mathcal{L})$. This shows that the space $PTh_\mathcal{L}=X^{\Lambda(X)}$ has implication. \cuad 

\begin{corollary}\label{distrissober}
A distributive space $X$ is homeomorphic to the sober space $X^{\Lambda(X)}$ with base $\Lambda(X^{\Lambda(X)})$ via the homeomorphism $x\mapsto x^{\Lambda(X)}$. \cuad
\end{corollary}

\begin{corollary} \label{distrspec2}
The bounded distributive spaces are exactly the spectral spaces. \cuad
\end{corollary}

\begin{corollary}\label{charspec}
Let $\mathcal{L}$ be an intuitionistic abstract logic. Then its space $X=PTh_\mathcal{L}$ is a spectral space with implication. \cuad
\end{corollary}

Since distributive spaces are sober (Corollary \ref{distrissober}), we call such spaces also \textbf{distributive sober spaces}, if we wish to emphasize the property of sobriety.

The following observation, whose proof is an easy exercise, establishes a close relationship between the topological properties of the distributive space $X$ and the algebraic properties of its base, the lattice of compact opens $\Lambda(X)$. The latter can be seen in some sense as an algebraic counterpart of the former. That is, we get an algebraic  characterization of the topological space $X$ by means of its base $\Lambda(X)$. 
 
\begin{lemma}\label{latticerepr}
Let $X$ be a distributive sober space.
\begin{enumerate}
\item $X$ is a spectral space with implication $\rightarrow$ if and only if $(\Lambda(X),\cup,\cap,\rightarrow)$ is a Heyting algebra.
\item $X$ is a boolean space with implication $\rightarrow$ if and only if $\Lambda(X)$ with $\rightarrow$ and the usual set-theoretic operations is a Heyting algebra that specializes to a boolean lattice.
\end{enumerate}
\cuad
\end{lemma}


\section{Stable logic maps}\label{stabil}

\hp So far we have studied the objects of the categories which will be defined in the next section. Let us determine the corresponding morphisms. In the case of spectral spaces these are, as expected, the spectral maps. In the larger category of distributive spaces we may also work with spectral maps, since the bases of these sober spaces are again sets of compact opens. For the morphisms between distributive logics we consider logic maps as studied in \cite{lew1}. We will need here only those logic maps whose pre  images preserve the prime theories. We call such logic maps \textit{stable}. 

\begin{definition}\label{stablelogmap}
Let $\mathcal{L},\mathcal{L'}$ be distributive abstract logics. A logic map is a function $h:Expr_\mathcal{L}\rightarrow Expr_\mathcal{L'}$ satisfying $\{h^{-1}(T')\mid T'\in Th_\mathcal{L'}\}\subseteq Th_\mathcal{L}$. We write $h:\mathcal{L}\rightarrow\mathcal{L'}$. A logic map $h$ is called stable if $\{h^{-1}(T')\mid T'\in PTh_\mathcal{L'}\}\subseteq PTh_\mathcal{L}$.\footnote{Since  $PTh_\mathcal{L}$ ($PTh_\mathcal{L'}$) is a generator set for $\mathcal{L}$ (for $\mathcal{L'}$), this condition implies the weaker condition $\{h^{-1}(T')\mid T'\in Th_\mathcal{L'}\}\subseteq Th_\mathcal{L}$.} A logic map $h$ is called normal if $\{h^{-1}(T')\mid T'\in Th_\mathcal{L'}\}=Th_\mathcal{L}$. 
\end{definition}

\begin{lemma}\label{normalstable}
Let $\mathcal{L},\mathcal{L'}$ be distributive abstract logics and let $h:Expr_\mathcal{L}\rightarrow Expr_\mathcal{L'}$ be any function. If $\{h^{-1}(T')\mid T'\in PTh_\mathcal{L'}\}= PTh_\mathcal{L}$, then $h$ is a normal and stable logic map.
\end{lemma}

\paragraph*{Proof.}
Suppose the premises hold. Let $T'\in Th_\mathcal{L'}$. Since $PTh_\mathcal{L'}$ is a generator set we have $T'=\bigcap\mathcal{T'}$ for some $\mathcal{T'}\subseteq PTh_\mathcal{L'}$. It follows that $h^{-1}(\bigcap\mathcal{T'})=\bigcap \{h^{-1}(T')\mid T'\in\mathcal{T'}\}\in Th_\mathcal{L}$. Hence, $h$ is a logic map. Now observe that $h$ is stable by hypothesis. We show that $h$ is normal. Let $T\in Th_\mathcal{L}$. Since $PTh_\mathcal{L}$ is a generator set, there is $\mathcal{T}\subseteq PTh_\mathcal{L}$ with $T=\bigcap\mathcal{T}$. Let $\mathcal{T'}:=\{T'\in PTh_\mathcal{L'}\mid h^{-1}(T')\in\mathcal{T}\}$. By hypothesis, this set is non-empty if $\mathcal{T}$ is non-empty. It follows that $h^{-1}(\bigcap\mathcal{T'})=\bigcap \{h^{-1}(T')\mid T'\in\mathcal{T'}\}=T$. Thus, $h$ is normal. $\cuad$

Recall that $=_\mathcal{L}$ denotes the relation of logical equivalence in logic $\mathcal{L}$.

\begin{lemma}
A logic map $h:\mathcal{L}\rightarrow\mathcal{L'}$ between distributive logics is stable iff $h(a\vee b)=_\mathcal{L'} h(a)\vee' h(b)$, for all $a,b\in Expr_\mathcal{L}$ and the respective connectives of disjunction of $\mathcal{L}$ and $\mathcal{L'}$.
\end{lemma}

\paragraph*{Proof.}
Suppose $h$ is stable and let $h(a\vee b)\in P'$ for any $P'\in PTh_\mathcal{L'}$. Then $a\vee b\in P=h^{-1}(P')$. Since $P$ is prime, $a\in P$ or $b\in P$. Thus, $h(a)\in P'$ or $h(b)\in P'$. Similarly for the other direction. Since $P'$ was arbitrarily chosen and the collection of all prime theories forms a generator set, it follows that $h$ preserves disjunction in the sense of the Lemma. Now suppose that $h$ preserves disjunction. Let $P'\in PTh_\mathcal{L'}$. $T=h^{-1}(P')$ is a theory. Let $a\vee b\in T$. Suppose $a\notin T$. Thus, $h(a)\notin P'$. Then $h(a\vee b)\in P'$ implies $h(b)\in P'$, that is, $b\in T$ and $T$ is prime. $\cuad$

\begin{remark}
In \cite{lew1} it is shown that the well-known G\"odel-translation $g:\mathcal{L}_{cl}\rightarrow\mathcal{L}_{int}$ from classical to intuitionistic propositional logic is a logic map (see Example 4 in \cite{lew1}). Recall that $g$ is defined as follows:
\begin{itemize}
\item $g(p)=\sim\sim p$, where $p$ is a propositional variable
\item $g(\sim a)=\sim g(a)$
\item $g(a\vee b)=\sim(\sim g(a)\wedge \sim g(b))$
\item $g(a\wedge b)=g(a)\wedge g(b)$
\item $g(a\rightarrow b)=g(a)\rightarrow g(b)$
\end{itemize}
Now observe that $g(p\vee q)=\sim(\sim\sim\sim p\wedge \sim\sim\sim q)=_\mathcal{L'}\sim(\sim p\wedge\sim q)\neq_\mathcal{L'}\sim\sim p\vee\sim\sim q=g(p)\vee g(q)$, for propositional variables $p,q$. By the preceding Lemma, $g$ cannot be stable.
\end{remark} 

In \cite{lew1} a \textit{logic isomorphism} from $\mathcal{L}$ to $\mathcal{L'}$ is given as a $L$-surjective normal logic map. In the same paper it is shown that this notion is equivalent with the concept of \textit{equipollence between logical systems} introduced and studied by Caleiro and Gon\c calves \cite{calgon}. We adopt here the notion of logic isomorphism. 

\begin{definition}
Let $\mathcal{L},\mathcal{L'}$ be distributive abstract logics and let $h:Expr_\mathcal{L}\rightarrow Expr_\mathcal{L'}$ be a logic map. $h$ is said to be $L$-surjective if for every $a'\in Expr_\mathcal{L'}$ there is some $a\in Expr_\mathcal{L}$ such that $h(a)=_\mathcal{L'} a'$. $h$ is called a logic isomorphism if $h$ is normal and $L$-surjective.
\end{definition}

\begin{remark}\label{logmaps}
\begin{itemize}
\item Example 5 in \cite{lew1} presents a logic map $i:\mathcal{L}_{int}\rightarrow\mathcal{L}_{cl}$ (the identity on the set of expressions) from intuitionistic to classical propositional logic, which is not normal. Nevertheless, $i$ is a stable logic map, since $i^{-1}=i$ maps a maximal (=prime) theory of $\mathcal{L}_{cl}$ to a maximal theory of $\mathcal{L}_{int}$.
\item If $h:\mathcal{L}\rightarrow\mathcal{L'}$ is a logic isomorphism, then there is a logic isomorphism $g:\mathcal{L'}\rightarrow\mathcal{L}$ such that $g(h(a))=_\mathcal{L} a$ and $h(g(a'))=_\mathcal{L'} a'$, for all $a\in Expr_\mathcal{L}$ and for all $a'\in Expr_\mathcal{L'}$. $g$ can be defined by $a'\mapsto a$ iff $h(a)=_\mathcal{L'} a'$ (see Theorem 4.15 \cite{lew1}). If $h_1:\mathcal{L}\rightarrow\mathcal{L'}$ and $h_2:\mathcal{L'}\rightarrow\mathcal{L''}$ are logic isomorphisms, then there is a logic isomorphism $h_3:\mathcal{L}\rightarrow\mathcal{L''}$. $h_3$ can be defined by $a\mapsto h_2(h_1(a))$ (see Theorem 4.16 \cite{lew1}).
\item If $h:\mathcal{L}\rightarrow\mathcal{L'}$ is a logic map and $a=_\mathcal{L} b$, then $h(a)=_\mathcal{L'} h(b)$ (see Proposition 3.2 \cite{lew1}).   
\end{itemize}
\end{remark}

Let $\mathcal{L}$ be a distributive abstract logic. For a formula $a\in Expr_\mathcal{L}$ we denote the equivalence class of $a$ modulo $=_\mathcal{L}$ by $\overline{a}$. A logic map $h:\mathcal{L}\rightarrow\mathcal{L'}$, $a\mapsto h(a)$, induces a function $h_*: \mathcal{L} / =_\mathcal{L} \longrightarrow  \mathcal{L'} / =_\mathcal{L'},\overline{a}\mapsto \overline{h(a)}$. By the last item of Remark \ref{logmaps} this function $h_*$ is well defined. We call it the map induced by $h$ in {\sl passing to the quotient}. We may identify $h_*$ with $h$ itself. So in the following, we identify formulas $a$ with their equivalence classes $\overline{a}$.

\section{Duality between the categories of intuitionistic abstract logics and spectral spaces with implication}\label{equivalence}

\hp In this section, we will establish the duality between the categories of intuitionistic abstract logics $\B{IL}$ and spectral spaces with implication $\B{SI}$. These two categories have on the one side, intuitionistic abstract logics as objects and stable logic maps as morphisms. On the other side, we have spectral spaces with implication as objects and spectral maps as morphisms.. 

The notion of the \textit{inverse complement} $G$ of a logic map $h:\mathcal{L}\rightarrow\mathcal{L'}$ is defined in \cite{lew1} where it is also shown that $G$ is a continuous map between the respective theory spaces. Also a condition is established, within the framework of abstract logics, which has the same form as the \textit{satisfaction condition of institutions} (see, e.g., \cite{gogbur}). In the present context, the inverse complement will play a similar role.

\begin{definition} \label{logmap}
Let $\mathcal{L}, \mathcal{L'}$ be minimally generated logics and let $h:\mathcal{L} \rightarrow \mathcal{L'}$ be a (stable) logic map. The {\sl inverse complement of $h$} is the map $G: Th_\mathcal{L'}\rightarrow Th_\mathcal{L}$ defined by: $G(T'):= h^{-1}(T')$.
\end{definition}

\begin{notation}
Denote by $\B{IL}$ the category whose objects are intuitionistic abstract logics and whose morphisms are stable logic maps. Denote by $\B{SI}$ the category whose objects are spectral spaces with implication and whose morphisms are spectral maps. Remark that it is not difficult to show that these are in fact categories. We omit the details.
\end{notation}

\nhp In a first step, we define the following contravariant functor \\ 
\hfl $\mathcal{F}: \B{IL} \longrightarrow \B{SI}$ \hfl  

  \nhp    \hfl    $ ob(\B{IL}) \ni \mathcal{L} \longmapsto \mathcal{F}(\mathcal{L}) := PTh_\mathcal{L} \in ob(\B{SI})\hfl \\
       \hfl  mor_{\B{IL}}(\mathcal{L}; \mathcal{L'}) \ni h \longmapsto \mathcal{F}(h)  \in mor_{\B{SI}}(PTh_\mathcal{L'}; PTh_\mathcal{L})$ \hfl \\
          defined by $\mathcal{F}(h): PTh_\mathcal{L'} \longrightarrow PTh_\mathcal{L},  P' \mapsto G(P'),$ 
             with $G$ the inverse complement of $h$.  

Note that the functor $\mathcal{F}$ is well-defined. By Corollary \ref{charspec}, $\mathcal{F}(\mathcal{L}):=PTh_\mathcal{L}$ with the given topology is a spectral space with implication. On the other hand, since $h$ is a stable logic map, $\mathcal{F}(h)(P') := G(P')=h^{-1}(P')$ is a prime theory.    
                            
\begin{proposition}
With the above notation, $\mathcal{F}(h)=G$ is a spectral map.
\end{proposition}              
 
\paragraph*{Proof:} Since the basic opens are precisely the compact opens, it suffices to show that $\mathcal{F}(h)^{-1}=G^{-1}$ maps a basic open to a basic open. We follow a similar argumentation as in \cite{lew1} where it was shown that the inverse complement is a continuous map between respective theory spaces. Let $U$ be a basic open in $PTh_\mathcal{L}$. Observe that $\mathcal{F}(h)^{-1}(U) = G^{-1}(U)$ and that $U=a^{PTh_{\mathcal{L}'}}$ for some $a \in Expr_\mathcal{L'}$. Then \\
 $P' \in G^{-1}(a^{PTh_{\mathcal{L}}}) \hem $ iff $\hem G(P') = h^{-1}(P') \in a^{PTh_\mathcal{L}} \hem $ 
  iff $\hem a \in h^{-1}(P') \hem $  \hem
  iff \hem $ P' \in h(a)^{PTh_{\mathcal{L}'}},$ thus $G^{-1}(a^{PTh_\mathcal{L}}) = h(a)^{PTh_\mathcal{L'}}$. This is again a basic (and compact) open. \cuad \\
  
 \nhp In a second step, we define the following contravariant functor \\ 
\hfl $\mathcal{G}: \B{SI} \longrightarrow \B{IL}$ \hfl 

    $ob(\B{SI}) \ni X \longmapsto \mathcal{G}(X) = \mathcal{L}\in ob(\B{IL})$\\ 
where $\mathcal{L}:=(\Lambda(X),Th_\mathcal{L},\{\cap,\cup,\rightarrow,\sim\}$ is given as in the proof of Theorem \ref{distrspec}\\
       \hfl  $mor_{\B{SI}}(X; X') \ni f \longmapsto \mathcal{G}(f)  \in mor_{\B{IL}}(\mathcal{G}(X'); \mathcal{G}(X))$ \hfl \\
          defined by $\mathcal{G}(f): \Lambda(X') \rightarrow \Lambda(X),  U' \mapsto \mathcal{G}(f)(U') := f^{-1}(U')$.  
          
In the proof of Theorem \ref{distrspec} it is shown that $\mathcal{G}(X)=\mathcal{L}:=(\Lambda(X),Th_\mathcal{L},\{\cap,\cup,\rightarrow,\sim\}$ is in fact an abstract intuitionistic logic. Since $f$ is a spectral map, the application $\mathcal{G}(f)$ is also well defined.
 
\begin{proposition}\label{stablemap}
With the above notation, $\mathcal{G}(f):\mathcal{G}(X')\rightarrow\mathcal{G}(X)$ is a stable logic map.
\end{proposition}              
 
\paragraph*{Proof:} Put $h:=\mathcal{G}(f)$. Note that for $U\in\Lambda(X)$, $h^{-1}(U)=\{U'\in\Lambda(X')\mid h(U')=U\}=\{U'\in\Lambda(X')\mid f^{-1}(U')=U\}$. The prime theories of $\mathcal{L'}$ (of $\mathcal{L}$) are precisely the prime filters on the lattice $\Lambda(X')$ (on $\Lambda(X)$), respectively. So it suffices to show that for any prime filter $P\subseteq\Lambda(X)$, $h^{-1}(P)$ is a prime filter on $\Lambda(X')$. Let $P\subseteq\Lambda(X)$ be a prime filter. Since $X$ is a distributive space, $P=x^{\Lambda(X)}$ for some $x\in X$. We have $h^{-1}(P)=\{h^{-1}(U)\mid U\in P\}=\{h^{-1}(U)\mid x\in U\}=\{U'\in\Lambda(X')\mid x\in f^{-1}(U')\}=\{U'\in\Lambda(X')\mid f(x)\in U'\}=f(x)^{\Lambda(X)}=:P'$, which is a prime filter on $\Lambda(X')$. \quad

\begin{definition}
The natural isomorphism for the objects $\mathcal{L} \in ob(\B{IL})$

\nhp \hfl $ \mathcal{L} \longrightarrow \mathcal{G}(\mathcal{F}(\mathcal{L}))$ is given by the function \hfl \\
\hfl $ a \mapsto \tau_\mathcal{L}(a) := a^{PTh_\mathcal{L}}, a\in Expr_\mathcal{L}$. \hfl

\end{definition}

For any $\mathcal{L} \in ob(\B{IL})$, the function $\tau_\mathcal{L}$ is in effect a logic isomorphism $\tau_\mathcal{L}:\mathcal{L}\rightarrow\mathcal{G}(\mathcal{F}(\mathcal{L}))$ as the following result shows.

\begin{theorem} \label{theoremintutionistic}
Every intuitionistic abstract logic $\mathcal{L}$ is isomorphic to the intuitionistic abstract logic $\mathcal{G}(\mathcal{F}(\mathcal{L}))$ via the logic isomorphism $\tau_\mathcal{L}$, $a\mapsto a^{PTh_\mathcal{L}}$. That is, $\mathcal{G} \circ \mathcal{F} = \B{1}_{\B {IL}}$ and the following diagramm commutes. \\
 \hfl $\bcosq{$\mathcal{L}$}{$\mathcal{G}(\mathcal{F}(\mathcal{L}))$}{$\mathcal{G}(\mathcal{F}(\mathcal{L'}))$}{$\mathcal{L'}$}{$\tau_\mathcal{L}$}{$\tau_\mathcal{L'}$}{$h$}{$\mathcal{G}(\mathcal{F}(h))$}$  \hfl

\end{theorem}

\paragraph*{Proof:}
Let $\mathcal{L}$ be an intuitionistic abstract logic. Corollary \ref{charspec} yields the implicative spectral space $\mathcal{F}(\mathcal{L})=X=PTh_\mathcal{L}$ with $\Lambda(X)=\{a^{PTh_\mathcal{L}}\mid a\in Expr_\mathcal{L}\}$ as base of compact opens. Recall that by Definition \ref{ordspec}, for $P\in X$,  $P^{\Lambda(X)}=\{a^{PTh_\mathcal{L}}\in \Lambda(X)\mid P\in a^{PTh_\mathcal{L}}\}=\{a^{PTh_\mathcal{L}}\in \Lambda(X)\mid a\in P\}$. Furthermore, $X^{\Lambda(X)}=\{P^{\Lambda(X)}\mid P\in X\}$ is the set of all prime filters on the lattice $\Lambda(X)$. The proof of Theorem \ref{distrspec} yields an abstract intuitionistic logic  
\begin{equation*}
\mathcal{G}(\mathcal{F}(\mathcal{L}))=\mathcal{L'}=(\Lambda(X),Th_\mathcal{L'},\{\cap,\cup,\rightarrow,\sim\}),
\end{equation*}
where $Th_\mathcal{L'}=\{\bigcap A\mid A\subseteq X^{\Lambda(X)}$, $A\neq\varnothing\}$, $\rightarrow$ is the implication of $X$ and $\sim a^{PTh_\mathcal{L}}:=a^{PTh_\mathcal{L}}\rightarrow\varnothing$. Note that $PTh_\mathcal{L'}=X^{\Lambda(X)}$ is the set of prime theories of $\mathcal{L'}$. Let us show that $\tau_\mathcal{L}:\mathcal{L}\rightarrow\mathcal{L'}$ is a logic isomorphism. For any $P\in PTh_\mathcal{L}$, $\tau_\mathcal{L}(P)=\{\tau_\mathcal{L}(a)\mid a\in P\}=P^{\Lambda(X)}$ and $\tau_\mathcal{L}^{-1}(P^{\Lambda(X)})=\{\tau_\mathcal{L}^{-1}(a^{PTh_\mathcal{L}})\mid a\in P\}=P$. Hence, $\{\tau_\mathcal{L}^{-1}(P')\mid P'\in PTh_\mathcal{L'}\}=\{\tau_\mathcal{L}^{-1}(P^{\Lambda(X)})\mid P\in X\}=\{P\mid P\in X\}=X=PTh_\mathcal{L}$. By Lemma \ref{normalstable}, $\tau_\mathcal{L}$ is a normal and stable logic map. Of course, $\tau_\mathcal{L}$ is $L$-surjective --- that is, $\tau_\mathcal{L}$ is surjective if it is viewed as the induced map which is defined on the 
 quotient modulo logical equivalence. Then by definition, $\tau_\mathcal{L}$ is a logic isomorphism. 

Finally, for $a \in  Expr_\mathcal{L}$ we get: $(\mathcal{G}(\mathcal{F}(h)) \circ\tau_\mathcal{L})(a)=(\mathcal{G}(G)\circ\tau_\mathcal{L})(a)=G^{-1}(\tau_\mathcal{L}(a)) = G^{-1}(a^{PTh_\mathcal{L}})=h(a)^{PTh_\mathcal{L'}}=\tau_\mathcal{L'}(h(a))=(\tau_\mathcal{L'}\circ h)(a)$, showing that the above diagramm commutes. \cuad   

\begin{definition}
The natural isomorphisms $\sigma_X : X \longrightarrow \mathcal{F}(\mathcal{G}(X))$ for the objects $X \in ob(\B{SI})$ is defined by $x \mapsto \sigma_X(x) := x^{\Lambda(X)}$.
\end{definition}

The preceding definition is justified by the next result.

\begin{theorem}
With the above notations, $\mathcal{F} \circ \mathcal{G} = \B{1}_{\B {SI}}$ and the following diagramm commutes. \\
\hfl $\bcosq{$X$}{$\mathcal{F}(\mathcal{G}(X))$}{$\mathcal{F}(\mathcal{G}(X'))$}{$X'$}{$\sigma_X$}{$\sigma_{X'}$}{$f$}{$\mathcal{F}(\mathcal{G}(f))$}$  \hfl

\end{theorem}

\paragraph*{Proof:} By Proposition \ref{specisdistr}, a spectral space $X$ is a distributive space. Theorem \ref{distrspec} now says that $\sigma_X$ given by $x\mapsto x^{\Lambda(X)}$ is an homeomorphism from the space $X$ to the spectral space $X^{\Lambda(X)}=\mathcal{F}(\mathcal{G}(X))$. It remains to show that the above diagramm commutes. For this let $f: X \rightarrow X'$ be a spectral map. By Proposition \ref{stablemap}, $h:=\mathcal{G}(f)=f^{-1}$ is a stable logic map $h:\mathcal{G}(X')\rightarrow\mathcal{G}(X)$ given by $U'\mapsto f^{-1}(U')=U\in\Lambda(X)$, for $U'\in\Lambda(X')$. Let $G$ be the inverse complement of $h$. In the proof of Proposition \ref{stablemap} we have seen that $G(x^{\Lambda(X)})=h^{-1}(x^{\Lambda(X)})=f(x)^{\Lambda(X')}$ for any $x\in X$. So we get $(\sigma_{X'}\circ f)(x)=\sigma_{X'}(f(x))=f(x)^{\Lambda(X')}=h^{-1}(x^{\Lambda(X)})=G(x^{\Lambda(X)})=\mathcal{F}(h)(x^{\Lambda(X)})=\mathcal{F}(h)(\sigma_X(x))=(\mathcal{F}(\mathcal{G}(f))\circ\sigma_X)(x)$. This shows that the diagramm commutes. \cuad

So the above show the following
 
\begin{theorem}\label{Equivalent}
The categories $\B{IL}$ and $\B{SI}$ are dually equivalent. \cuad
\end{theorem}

The category of \textit{distributive abstract logics} is given by distributive abstract logics as objects and stable logic maps as morphisms. The category of \textit{distributive sober spaces} is given by distributive sober spaces as objects and spectral maps as morphisms. Generalizing our preceding results in an obvious way we get the equivalence of these larger categories. 

\begin{corollary}\label{EquivDistr}
The category of distributive abstract logics and the category of distributive sober spaces are dually equivalent. \cuad
\end{corollary}

\end{document}